\documentclass[12pt]{article}
\usepackage{amsmath}
\usepackage{amsfonts}
\usepackage{amssymb}
\usepackage{amsthm}
\usepackage{cite}

\newcommand{\ph}{\phantom}
\newcommand{\rz}{r}
\newcommand{\tr}{\rho}
\newcommand{\teps}{\epsilon} 
\newcommand{\chir}{\chi(\mco_{\rz})}
\newcommand{\chitr}{\chi(\mco_{\tr})}
\newcommand{\chitrz}{\chi_{\vz}(\mco_{\tr})}
\newcommand{\tchi}{\wt{\chi}}

\newcommand{\nor}{|\!|\!|}

\newcommand{\omar}{\om^{-\alr}}
\newcommand{\alr}{\gamma}
\newcommand{\te}{\hat{e}}
\newcommand{\cE}{\sup_{|\vep|\leq E}|\tilh(\vep)|^{-1}}

\newcommand{\cErs}{\sup_{|\vep|\leq E}|\tilh(\vep)|^{-2}}
\newcommand{\bsharp}{\mbox{\boldmath $^\sharp$}}
\newcommand{\bnatural}{\mbox{\boldmath $^\natural$}}
\newcommand{\hz}{h}
\newcommand{\tilh}{\widetilde{h}}
\newcommand{\Q}{Q}

\newcommand{\wt}{\widetilde}

\newcommand{\wFz}{\widetilde{F}_{\vz}}
\newcommand{\tF}{\widetilde{F}}

\newcommand{\Fpm}{F^{\pm}}
\newcommand{\cs}{ 2^{\fr{s}{2}-2}\Ga(\fr{s}{2}-1)}
\newcommand{\h}{\half}

\newcommand{\czn}{C_0^{\infty}(\real^s)}

\newcommand{\dsp}{d^sp_1\ldots d^sp_n}
\newcommand{\ds}{d^sp_2\ldots d^sp_n}
\newcommand{\pp}{\vep_2,\ldots,\vep_n}
\newcommand{\tmn}{\tau_{\mub,\nub}}
\newcommand{\p}{*}

\newcommand{\reals}{\real^{s+1}}

\newcommand{\x}{x_1\ldots x_N}

\newcommand{\pin}{\fr{1}{p}}

\newcommand{\fpm}{f^{\pm}}

\newcommand{\omp}{\om^{\h}}
\newcommand{\omm}{\om^{-\h}}

\newcommand{\om}{\omega}

\newcommand{\norm}{|\!|\!|\cdot |\!|\!|}
\newcommand{\lin}{\mathcal{L}}

\newcommand{\vy}{\vec{y}}

\newcommand{\PiE}{\Pi_E}
\newcommand{\al}{\alpha}
\newcommand{\be}{\beta}

\newcommand{\la}{\lambda}
\newcommand{\vp}{\varphi}
\newcommand{\Ga}{\Gamma}

\newcommand{\eps}{\varepsilon}

\newcommand{\de}{\delta}

\newcommand{\TEp}{T_{E,+}}
\newcommand{\TEm}{T_{E,-}}
\newcommand{\TEpm}{T_{E,\pm}}

\newcommand{\Tbpm}{T_{h,\pm}}
\newcommand{\Tbp}{T_{h,+}}
\newcommand{\Tbm}{T_{h,-}}
\newcommand{\LL}{\mathcal{L}}
\newcommand{\Lp}{\mathcal{L}^{+}}
\newcommand{\Lm}{\mathcal{L}^{-}}
\newcommand{\Lpm}{\mathcal{L}^{\pm}}
\newcommand{\LJ}{\mathcal{L}}

\newcommand{\nat}{\mathbb{N}}

\newcommand{\real}{\mathbb{R}}

\newcommand{\mup}{\mu^+}
\newcommand{\mum}{\mu^-}
\newcommand{\nup}{\nu^+}
\newcommand{\num}{\nu^-}

\newcommand{\mub}{\overline{\mu}}
\newcommand{\nub}{\overline{\nu}}
\newcommand{\nuh}{\nub_a}
\newcommand{\nuc}{\nub_b}

\newcommand{\Smn}{S_{\mub,\nub}}
\newcommand{\trace}{\mathcal{T}}
\newcommand{\traceE}{\trace_E}

\newcommand{\traceEB}{\trace_{E,1}}

\newcommand{\trac}{\mathring{\trace}_{E}}

\newcommand{\vxb}{\underline{x}}
\newcommand{\vep}{\vec{p}}

\newcommand{\veq}{\vec{q}}

\newcommand{\vz}{\vec{x}}

\newcommand{\vx}{\vec{x}}
\newcommand{\Nnat}{N_{\bnatural}}

\newcommand{\Ns}{N_{\bsharp}}

\newcommand{\fun}{\vp}

\newcommand{\hil}{\mathcal{H}}
\newcommand{\mfa}{\mathfrak{A}}
\newcommand{\mco}{\mathcal{O}}

\newcommand{\fr}[2]{\frac{#1}{#2}}

\newcommand{\non}{\nonumber}

\newcommand{\vac}{\Omega}
\newcommand{\fp}{f^+}
\newcommand{\fm}{f^-}
\newcommand{\half}{\fr{1}{2}}
\newcommand{\lan}{\langle}
\newcommand{\ran}{\rangle}

\def\proof{\noindent{\bf Proof. }}
\def\qed{$\Box$\medskip}

\newtheorem{theoreme}{Theorem } [section]
\newtheorem{proposition}[theoreme]{Proposition}
\newtheorem{lemma}[theoreme]{Lemma}
\newtheorem{definition}[theoreme]{Definition}
\newtheorem{corollary}[theoreme]{Corollary}
\newtheorem{remark}[theoreme]{Remark}
\newtheorem{example}[theoreme]{Example}
\newtheorem{criterion}[theoreme]{Criterion}

\newcommand{\beq}{\begin{equation}}
\newcommand{\eeq}{\end{equation}}
\newcommand{\beqa}{\begin{eqnarray}}
\newcommand{\eeqa}{\end{eqnarray}}
\newcommand{\ben}{\begin{arabicenumerate}}
\newcommand{\een}{\end{arabicenumerate}}
\newcommand{\bex}{\begin{example}}
\newcommand{\eex}{\end{example}}
\newcommand{\ber}{\begin{remark}}
\newcommand{\eer}{\end{remark}}
\newcommand{\bec}{\begin{corollary}}
\newcommand{\eec}{\end{corollary}}
\newcommand{\bep}{\begin{proposition}}
\newcommand{\eep}{\end{proposition}}
\newcommand{\becr}{\begin{criterion}}
\newcommand{\eecr}{\end{criterion}}

\def\bel{\begin{lemma}}
\def\eel{\end{lemma}}
\def\bet{\begin{theoreme}}
\def\eet{\end{theoreme}}
\def\bed{\begin{definition}}
\def\eed{\end{definition}}

\begin{document}

\title{A Sharpened Nuclearity Condition for Massless Fields}
\author{Wojciech Dybalski\\ [5mm] 
  Institut f\"ur Theoretische Physik, Universit\"at G\"ottingen, \\[2mm]
Friedrich-Hund-Platz 1, D-37077 G\"ottingen - Germany \\ [2mm]
 e-mail: dybalski@theorie.physik.uni-goettingen.de}
\date{}
\maketitle
\begin{abstract}
A recently proposed phase space condition which comprises information about the vacuum structure and timelike
asymptotic behavior of physical states is verified in massless free field theory. There follow interesting conclusions
about the momentum transfer of local operators in this model.
\end{abstract}

\section{Introduction}
Compactness and nuclearity conditions, which characterize phase space properties, proved useful in the study of many aspects of Quantum Field Theory \cite{HS, BW, BJ, Scaling, Bos1, Bos2, Gan1}. Verification of phase space conditions in models \cite{BW,BJ1,universal,BP, Bos1, Dyb} is an integral part of these  investigations, since it demonstrates consistency of these criteria with the basic postulates of local, relativistic quantum physics \cite{Haag}. In \cite{Dyb} a sharpened nuclearity condition has been proposed. It restricts correlations between different phase space regions and implies several physically desirable features. Among them are a certain form of additivity of energy over isolated subsystems and the uniqueness of vacuum states which can be prepared with a finite amount of energy. These  vacuum states appear, in particular,  as  limits of physical states under large timelike translations in Lorentz covariant theories and are approximated by states of increasingly sharp energy-momentum values, in accordance with the uncertainty principle. This novel nuclearity condition seems also relevant to the study of particle aspects of a theory \cite{BS}. It is the aim of the present Letter to verify this criterion in massless free field theory. In comparison with the
massive case studied in \cite{Dyb}, the present investigation requires substantial technical improvements which we discuss below. As  will be shown in a future publication, these advances enable a detailed harmonic analysis of translation automorphisms in massless theories.

Before we formulate the sharpened nuclearity condition, we recall briefly the mathematical framework: Let $V$, $W$ be Banach spaces 
and $\norm$ be a norm on the space $\lin(V,W)$ of linear maps from $V$ to $W$. We say that a map $\Pi: V\to W$ is p-nuclear w.r.t. the norm $\norm$ if there exists a decomposition $\Pi(v)=\sum_n\Pi_n(v)$ into rank-one maps, convergent for any $v\in V$ in the norm topology in $W$, s.t. $\nu:=(\sum_n\nor\Pi_n\nor^p)^{\fr{1}{p}}<\infty$. The $p$-norm $\nor\Pi\nor_{p}$ of this map is the smallest such $\nu$ over the set of all admissible decompositions. To construct
the norms which are suitable for our purposes, suppose that there acts a group of automorphisms $\real^{s+1}\ni x\to\be_x$ on $V$. Then, for any $N\in\nat$ and $x_1\ldots x_N\in\real^{s+1}$, we set
\beq
\|\Pi\|_{\x}=\sup_{v\in V_1}\bigg(\sum_{k=1}^N\|\Pi(\be_{x_k}v)\|^2\bigg)^{\half},\quad \Pi\in \lin(V,W), \label{Nnorm}
\eeq
where $V_1$ is the unit ball in $V$, and denote the corresponding $p$-norm by $\|\cdot\|_{p,\x}$.

Next, we identify the spaces $V$, $W$, automorphisms $\be_x$ and  maps $\Pi$ in the framework of Quantum Field Theory.
Let $\hil$ be the Hilbert space,
$\om_0$ the normal vacuum state, $\real^{s+1}\ni x\to\al_x\in \textrm{Aut}(B(\hil))$ the translation automorphisms and $H$ the Hamiltonian. We set $\traceE=P_E B(\hil)_*P_E$, where $P_E$ is the spectral projection of $H$ on the subspace 
spanned by vectors of energy lower than $E$ and  choose $V=\trac:=\{\fun-\fun(I)\om_0 \ | \ \fun\in\traceE\}$.
This space is clearly invariant under the dual action of translations $\be_x=\al^\p_x$. Finally, we set
$W=\mfa(\mco)^*$, where $\mfa(\mco)\subset B(\hil)$ is the local algebra of observables attached to a double cone $\mco\subset \real^{s+1}$, and define the family of maps $\PiE: \trac\to \mfa(\mco)^*$ given by
\beq
\PiE(\fun)=\fun|_{\mfa(\mco)},\quad \fun\in\trac.
\eeq
The strengthened nuclearity condition, proposed in \cite{Dyb}, has the following form.
\begin{enumerate}
\item[] \bf Condition \rm $\Nnat$. The maps $\PiE$ are $p$-nuclear w.r.t. the norms $\|\cdot\|_{\x}$
for any $N\in\nat$, $\x\in\real^{s+1}$,  $0<p\leq 1$, $E\geq 0$,
and double cone $\mco\subset\reals$. Moreover, there holds for their nuclear $p$-norms
\beq
\limsup\|\PiE\|_{p,\x}\leq c_p, \label{strengthening}
\eeq
where $c_p$ is independent of $N$ and the limit is taken for configurations $\x$, where all $x_i-x_j$, $i\neq j$,
tend to spacelike infinity.
\end{enumerate}
We note that the first, qualitative part of this criterion is equivalent to Condition~$\Ns$ formulated in \cite{BP}
and the essential additional information is contained in the bound~(\ref{strengthening}).
This refinement is motivated by the observation that a measurement is always accompanied
by an energy transfer from the physical state to the observable. Additivity of energy over isolated 
subregions  should then imply that for any $\fun\in\trac$ the restricted functionals $\al_{\vx}^\p\fun|_{\mfa(\mco)}$ are arbitrarily close to zero apart from translations varying in some compact subset of $\real^{s}$, depending on $\fun$. This picture is particularly plausible in a massive theory, where a state of bounded energy contains only a finite number of particles which are well localized in space. Making use of this simplification,
Condition~$\Nnat$ was verified in \cite{Dyb} in a theory of non-interacting massive particles.

In the present Letter we demonstrate that this criterion is valid also in the massless case for 
$s\geq 3$. There the status of Condition $\Nnat$ is less obvious, since one has to handle the "infrared cloud"- states of bounded energy containing arbitrarily large numbers of massless particles whose localization properties are poor. The proof is accomplished by combining the underlying physical idea of additivity of energy  over isolated subregions (Lemma~\ref{harmonic}) with
the quadratic decay of vacuum correlations between spatially separated observables in a massless theory
(Lemma~\ref{Cook}). As an interesting application of our methods, we briefly discuss in the Conclusions the momentum transfer of local operators in the model under study.

\section{Massless Scalar Free Field Theory}

In the model at hand  the Hilbert space $\hil$ is the symmetric Fock space
over $L^2(\real^s, d^sp)$. On this latter space there acts the unitary representation
of translations 
\beq 
(U_1(x)f)(\vep)=e^{i(\om(\vep)x^0-\vep \vx)}f(\vep),\quad f\in L^2(\real^s, d^sp),
\eeq
where $\om(\vep)=|\vep|$. We denote by $U(x)$ its second quantization acting on $\hil$,
introduce the corresponding family of automorphisms of $B(\hil)$
\beq
\al_x(\cdot)=U(x)\cdot U(x)^*
\eeq
and adopt the notation $A(x):=\al_x(A)$ for translated operators $A\in B(\hil)$.
Next, we construct the local algebra $\mfa(\mco)$ attached to the double cone $\mco$,
whose base is the $s$-dimensional ball $\mco_{\rz}$ of radius $\rz$ centered at the origin in
configuration space: We introduce the closed subspaces $\Lpm:=[\om^{\mp\fr{1}{2}}\widetilde{D}(\mco_{\rz})]$, 
where tilde denotes the Fourier transform, represent the respective projections by the same symbol
and consider the real linear subspace of $L^2(\real^s, d^sp)$
\beq
\LJ=(1+J)\Lp+(1-J)\Lm,
\eeq
where $J$ is the complex conjugation in configuration space.
Then the local algebra is given by
\beqa
\mfa(\mco)=\{ \ W(f) \ | \ f\in\LJ \ \}^{\prime\prime},
\eeqa
where $W(f)=e^{i(a^*(f)+a(f))}$ and $a^*(f)$, $a(f)$ are the creation and annihilation operators.

The rest of this section, which serves mostly to establish our notation, is devoted to the proof
of the well known fact \cite{BP,Bos3} that the maps $\Pi_E$ in this model are $p$-nuclear w.r.t. 
the standard norm on $\lin(\trac,\mfa(\mco)^*)$. In the massive case the argument
was outlined in \cite{Dyb}, Appendix B, so it suffices here to give a brief sketch which stresses the modifications:
First, our present construction of the trace-class operator $T$ differs from the choices made in
the existing literature \cite{BP,Bos3,Dyb}:  Let $\Q_E$ be the projection on states of energy lower than $E$ in the single-particle space,
let $\hz\in D(\mco_{\rz})$ be real and s.t. $\tilh>0$. We choose $\fr{1}{2}\leq\alr<\fr{s-1}{2}$
and define operators $\TEpm=\omm \Q_E\Lpm$, $\Tbpm=\om^{-\alr} \tilh^{1/2}\Lpm$, where $\tilh$ is the corresponding multiplication operator in momentum space.
By a slight modification of  Lemma~3.5  from \cite{BP} one obtains that for $s\geq 3$ these operators satisfy
$\||\TEpm|^p\|_1<\infty$, $\||\Tbpm|^p\|_1<\infty$ for any $p>0$, where $\|\cdot\|_1$ denotes
the trace norm. We define the operator $T$ as follows
\beq
T=(|\TEp|^2+|\TEm|^2+|\Tbp|^2+|\Tbm|^2)^\h.
\eeq
Making use of the fact  \cite{Kos} that for any $0<p\leq1$ and any pair of positive operators $A$, $B$, s.t.
$A^p$, $B^p$ are trace-class, there holds $\|(A+B)^p\|_1\leq \|A^p\|_1+\|B^p\|_1$, we get
\beq
\|T^p\|_1\leq\||\TEp|^p\|_1+\||\TEm|^p\|_1 +\||\Tbp|^p\|_1+\||\Tbm|^p\|_1 \textrm{ for } 0<p\leq 1. \label{lub3}
\eeq
Since $T$ commutes with $J$, it has a $J$-invariant orthonormal basis of eigenvectors $\{e_j\}_1^\infty$ 
and we denote the corresponding eigenvalues by $\{t_j\}_1^\infty$.

In order to construct an expansion of the map $\PiE$ into rank-one mappings, we evaluate a Weyl operator on some functional $\fun\in\trac$, rewrite it in a normal ordered form and expand it into a power series
\beqa
&\fun&\!\!\!\!\!(W(f))\non\\
&=&e^{-\fr{1}{2}\|f\|^2}
\sum_{m^{\pm},n^{\pm}\in\nat_{0}}\fr{i^{m^++n^++2m^-}}{m^+!m^-!n^+!n^-!}
\fun(a^*(\fp)^{m^+}a^*(\fm)^{m^-}a(\fp)^{n^+}a(f^-)^{n^-}),\qquad \label{powerseries}
\eeqa
where $f=f^++if^-$ and $f^{\pm}\in\Lpm$ are real in configuration space.
Subsequently, we expand each function $f^\pm$ in the orthonormal basis $\{e_j\}_1^{\infty}$ of  $J$-invariant eigenvectors of the operator $T$: $f^\pm=\sum_{j=1}^{\infty}e_j\lan e_j|f^\pm\ran$. Then, making use of the multinomial formula, we obtain
\beq
a^{(*)}(f^\pm)^{m^\pm}=\sum_{\mu^\pm,|\mu^\pm|=m^\pm}\fr{m^\pm!}{\mu^\pm!}\lan e|\fpm\ran^{\mu^\pm} a^{(*)}(\Lpm e)^{\mu^\pm},
\label{multinomial}
\eeq
where $\mup$, $\mum$ are multiindices, and substitute these expansions to (\ref{powerseries}). In order to simplify
the resulting expression, we define for any two pairs of multiindices $\mub=(\mup,\mum)$, $\nub=(\nup,\num)$ functionals $\Smn\in\trac^*$ given by
\beq
\Smn(\fun)=\fun(a^*(\LL e)^{\mub}a(\LL e)^{\nub} ),
\eeq
where $a^{(*)}(\LL e)^{\mub}=a^{(*)}(\Lp e)^{\mup}a^{(*)}(\Lm e)^{\mum}$. Moreover, with the help of the formula
\beqa
(\vac|[a(e_1),[\ldots,[a(e_k),[a^*(e_{k+1}),[\ldots, [a^*(e_l),W(f)],\ldots]\vac)\non\\
=e^{-\half\|f\|^2}\prod_{n_1=1}^k\lan e_{n_1}| if\ran \prod_{n_2=k+1}^l\lan if| e_{n_2}\ran,
\eeqa
one can express the factors $\lan e|\fpm\ran^{\mu^\pm}$, appearing in (\ref{multinomial}), in terms of
normal functionals $\tau_{\mub,\nub}\in \mfa(\mco)^*$ defined
as in \cite{Dyb}, Appendix~B, (using methods from \cite{Bos3}).  Then expression~(\ref{powerseries})
takes the form
\beqa
\fun(W(f))&=&\sum_{\mub,\nub}\tau_{\mub,\nub}(W(f)) \Smn(\fun).  \label{ffexpansion}
\eeqa
In order to extend this formula to all $A\in \mfa(\mco)$, we study its convergence properties:
In the present case the norms of the functionals $\tau_{\mub,\nub}$
are not uniformly bounded in $\mub$, $\nub$. Instead, one  obtains as in  formula~(B.7) of \cite{Dyb}
\beq
\|\tau_{\mub,\nub}\|\leq \fr{4^{|\mub|+|\nub|} }{(\mub!\nub!)^\h} \bigg(\fr{(\mub+\nub)!}{\mub!\nub!}\bigg)^\h\leq 
\fr{2^{\fr{5}{2}(|\mub|+|\nub|)}}{(\mub!\nub!)^\h},\label{tauestimate}
\eeq 
where $|\mub|=|\mup|+|\mum|$ and $\mub!=\mup!\mum!$.
Making use of the fact that for any $f_1,\ldots,f_n\in L^2(\real^s, d^sp)$ in the domain of $\omp$ there hold the so called energy bounds~\cite{BP}
\beq
\|a(\omp f_1)\ldots a(\omp f_n)P_E\|
\leq (E)^{\fr{n}{2}}\|f_1\|\ldots \|f_n\|,
\eeq
we obtain the estimate
\beq
\|\Smn\|\leq E^{\fr{|\mub|+|\nub|}{2}}\|\omm\Q_E \LL e\|^{\mub}\,\|\omm\Q_E\LL e\|^{\nub} \leq E^{\fr{|\mub|+|\nub|}{2}}t^{\mub} t^{\nub}. \label{Sestimate}
\eeq 
With the help of the bounds (\ref{tauestimate}) and (\ref{Sestimate}) one verifies that for
any $0<p\leq 1$
\beqa
\sum_{\mub,\nub }\|\tmn\|^p \, \|\Smn\|^p \leq \sum_{\mub,\nub} 
\fr{(2^5E)^{\half p (|\mub|+|\nub|)} }{(\mub!)^{\half p}(\nub!)^{\half p}} t^{p\mub}t^{p\nub} 
&=& \bigg(\sum_{\mup}\fr{(2^5E)^{\half p |\mup|} }{(\mup!)^{\half p}} t^{p\mup}\bigg)^4\non\\
&\leq& \bigg(\sum_{k=0}^\infty \fr{(2^5E)^{\half pk}\|T^p\|_1^k}{(k!)^{\half p}} \bigg)^4\!\!\!,\label{traces}
\eeqa
where in the last step we set $k=|\mup|$ and made use of the multinomial formula.
This bound allows us to restate expression (\ref{ffexpansion}) as follows
\beqa
\PiE(\fun)&=&\sum_{\mub,\nub}\tau_{\mub,\nub}\Smn(\fun), \textrm{ for } \fun\in\trac, \label{exp1}
\eeqa
where the sum converges in the norm topology in $\mfa(\mco)^*$ and there holds, in addition,
$\|\PiE\|_{p}\leq (\sum_{\mub,\nub }\|\tmn\|^p \, \|\Smn\|^p)^{1/p}<\infty$ for $0<p\leq 1$. This concludes the proof
of the known fact that Condition $\Ns$ holds in massless free field theory \cite{BP,Bos3}.
In the next section we will use the same expansion (\ref{exp1}) to verify  Condition~$\Nnat$.
\section{Verification of Condition $\Nnat$}
By definition of the nuclear $p$-norms and formula (\ref{exp1}) there holds the bound
\beq
\|\PiE\|_{p,\x}\leq\bigg(\sum_{\mub,\nub}\|\tau_{\mub,\nub}\|^p\|\Smn\|^p_{\x}\bigg)^{\pin}.\label{start}
\eeq
To verify Condition $\Nnat$ we have to find estimates on
the norms $\|\Smn\|_{\x}$ whose growth with $N$ can be controlled at large spacelike distances $x_i-x_j$ 
for $i\neq j$.
The first step in this direction is taken in the following lemma which is inspired
by Lemma 2.2 from \cite{Buch3}. In contrast to the bound from \cite{Dyb}, Lemma~4.1,
the present estimate is uniform in the particle number and depends only on the energy of 
the state in question. This result substantiates the underlying physical idea of additivity 
of energy over isolated subregions.
\bel\label{harmonic} 
Suppose that $g \in L^2(\real^s, d^sp)$ and $\tilh g$ is in the domain of $\omm$, where $\tilh\in \widetilde{D}(\mco_{\rz})$
appeared in the definition of the operator $T$ above.
Then, for any  $x_1\ldots x_N\in\real^{s+1}$,  there holds the bound
\beqa
\|P_E\sum_{k=1}^N(a^*(g)a(g))(x_k)P_E\|\leq E\sup_{|\vep|\leq E}|\tilh(\vep)|^{-2}
\big\{ \| \omm\tilh g \|^2 \non\\
\phantom{4444444444}+(N-1)\sup_{i\neq j}|\lan \omm \tilh g|U(x_{i}-x_{j})\omm\tilh g\ran| \big\}.
\label{harmonice}
\eeqa
\eel
\proof 
We pick  single-particle vectors $\Psi_1, g_1\in L^2(\real^s,d^sp)$ and define  $Q=$\\ $\sum_{k=1}^N(a^*(g_1)a(g_1))(x_k)$. Then there holds
\beqa
(\Psi_1|QQ\Psi_1)\leq\sum_{l=1}^N(\Psi_1|(a^*(g_1)a(g_1))(x_l)\Psi_1)\sum_{k=1}^N|\lan U(x_k)g_1|U(x_l)g_1\ran|
\ph{4}& &\non\\
\leq (\Psi_1|Q\Psi_1)\big\{\|g_1\|^2+(N-1)\sup_{i\neq j}|\lan U(x_{j})g_1|U(x_{i})g_1\ran|\big\},& &
\eeqa
where we made use of the fact that $a(U(x_k)g_1)a(U(x_l)g_1)\Psi_1=0$ and of the Cauchy-Schwarz inequality.
Since $(\Psi_1|Q\Psi_1)^2\leq(\Psi_1| QQ \Psi_1)\|\Psi_1\|^2$, we obtain
\beqa
& &\sum_{k=1}^N(\Psi_1|(a^*(g_1)a(g_1))(x_k)\Psi_1)\non\\
& &\phantom{4444444}\leq \|\Psi_1\|^2\big\{\|g_1\|^2+(N-1)\sup_{i\neq j}|\lan U(x_{j})g_1|U(x_{i})g_1\ran|\big\}. \label{single}
\eeqa
Next, let $n\geq 1$ and $\Psi_n\in P_E\hil$ be an $n$-particle vector s.t. the corresponding symmetric wave-function 
$\Psi_n(\vep_1\ldots \vep_n)$ belongs to $S(\real^{s\times n})$. We also introduce a single-particle wave-function associated with
$\Psi_n$ given by  $\Psi_1(\vep_1)_{\pp}=|\vep_1|^\half \tilh(\vep_1)^{-1}\Psi_n(\vep_1,\ldots \vep_n)$, where we treat $\pp$ as parameters. With the help of (\ref{single}) we get
\beqa
\sum_{k=1}^N(\Psi_n|(a^*(g)a(g))(x_k)\Psi_n)\ph{444444444444444444444444444444444444444444}& &\non\\
=n\int\ds \sum_{k=1}^N(\Psi_{1,\pp}|( a^*(\omm\tilh g)a(\omm\tilh g) )(x_k)\Psi_{1,\pp})\ph{.}& &\non\\
 \leq n\int\dsp |\tilh(\vep_1)|^{-2} |\vep_1||\Psi_n(p_1,\ldots p_n)|^2\ph{444444444444444444444.}& &\non\\
\cdot\big\{\| \omm\tilh g \|^2+
(N-1)\sup_{i\neq j}|\lan \omm\tilh g|U(x_{i}-x_{j}) \omm\tilh g \ran|\big\}.& &
\eeqa
Finally, we note that
\beqa
& &n\int\dsp |\tilh(\vep_1)|^{-2} |\vep_1||\Psi_n(\vep_1,\ldots \vep_n)|^2\non\\
& &\phantom{4444}\leq \sup_{|\vep|\leq E}|\tilh(\vep)|^{-2}
\int\dsp (|\vep_1|+\cdots+|\vep_n|)|\Psi_n(\vep_1,\ldots \vep_n)|^2\non\\
& &\phantom{4444}\leq \sup_{|\vep|\leq E}|\tilh(\vep)|^{-2}  E\|\Psi_n\|^2\!,
\eeqa
where we made use of the fact that the wave-function is symmetric. Since the operators $(a^*(g)a(g))(x_k)$
conserve the particle number and vectors of the form $\Psi=c\vac+\sum_{n=1}^{\infty} \Psi_n$, where  $\|\Psi\|^2=|c|^2+\sum_{n=1}^{\infty}\|\Psi_n\|^2<\infty$,
are dense in $P_E\hil$, we easily obtain the bound in the statement of the lemma. \qed\\
Our next task is to control the expressions appearing on the right-hand side of estimate~(\ref{harmonice}).
Lemma \ref{F} below, which holds in particular for $\tF(\vep)=|\vep|^{-2}$,  will be crucial in this respect.  
We start with some definitions:
for any $\tr>0$ and some fixed $\teps>0$ we choose a function $\chi(\mco_{\tr})\in\czn$  s.t. $\chi(\mco_{\tr})(\vx)=1$ 
for $\vx\in\mco_{\tr}$ and $\chi(\mco_{\tr})(\vx)=0$ for $\vx\notin\mco_{\tr+\teps}$.
We denote the operator of multiplication by $\chi(\mco_{\tr})$ in configuration space by the same 
symbol.
\bel\label{F} Suppose that $F\in S^\prime(\real^s)$ coincides with a bounded, measurable function in the 
region $\{\, \vy\in\real^s \,|\, |\vy|\geq \tr\,\}$ and its Fourier transform
$\tF$ is a positive, measurable function s.t. $\tF^{1/2}\in L^2(\real^s,d^sp)+L^{\infty}(\real^s,d^sp)$.
Then $\tF^{1/2}\chitr$ is a bounded operator and there holds
\beq
\|\chitr \tF\chitrz\|\leq c_{s,\tr,\teps}\sup_{|\vec{z}|\leq 2\tr+3\teps} |F(\vec{z}-\vz)| \ \textrm{ for }  
|\vz|\geq 3(\tr+\teps),\label{Fbound}
\eeq
where $\chitrz(\vy)=\chitr(\vy-\vz)$, the constant $c_{s,\tr,\teps}$ is independent of $\vz$ and we denote the
operator of multiplication by $\tF$ in momentum space by the same symbol.
\eel
\proof In order to prove the first statement we make a decomposition $\tF^{1/2}=\tF^{1/2}_2+\tF^{1/2}_{\infty}$,
where $\tF^{1/2}_2\in L^2(\real^s,d^sp)$, $\tF^{1/2}_{\infty}\in L^{\infty}(\real^s,d^sp)$. Since 
$\tF^{1/2}_{\infty}$ is a bounded operator, it suffices to consider $\tF^{1/2}_2\chitr$. We pick
$f_1,f_2\in S(\real^s)$ and estimate
\beqa
|\lan f_1|\tF^{1/2}_2  \chitr f_2\ran|=(2\pi)^{-\fr{s}{2}}\big|\int d^spd^sq \ \bar{f}_1(\vep) \tF^{1/2}_2(\vep)\tchi(\mco_{\tr})(\vep-\veq)f_2(\veq)\big|\ph{4}& &\non\\
\leq c\|\bar{f}_1\tF^{1/2}_2\|_1\|\tchi(\mco_{\tr}) \|_2\|f_2\|_2\leq c\|f_1\|_2
\|\tF^{1/2}_2\|_2\|\tchi(\mco_{\tr}) \|_2\|f_2\|_2,& &
\eeqa
where in the second step we made use of the Young inequality\footnote 
{The Young inequality states that 
for any  positive functions $f\in L^{r_1}(\real^s,d^sp)$, $g\in L^{r_2}(\real^s,d^sp)$, $h\in L^{r_3}(\real^s,d^sp)$,
where 
$1\leq r_1,r_2,r_3\leq\infty$ s.t. $\fr{1}{r_1}+\fr{1}{r_2}+\fr{1}{r_3}=2$, there holds the
bound
\begin{eqnarray*}
\int d^spd^sq \ f(\vep)g(\vep-\veq)h(\veq)\leq c_{r_1,r_2,r_3}\|f\|_{r_1}\|g\|_{r_2}\|h\|_{r_3}.
\end{eqnarray*}} \cite{RS2} and in the last estimate we applied H\"older's inequality.

Next, we  verify relation (\ref{Fbound}). If $|\vz|\geq 3(\tr+\teps) $,
then $|\vy+\vz|\leq 2\tr+3\teps$ implies  $|\vy|\geq \tr$ and the  expression
\beq
\wFz(\vep):=(2\pi)^{-\fr{s}{2}}\int d^sy \, e^{-i\vep\vy}F(\vy)\chi_{-\vz}(\mco_{2(\tr+\teps)})(\vy) \label{Fx}
\eeq
defines a bounded, continuous function.
The operator of multiplication by $\wFz$ in momentum space, denoted by the same symbol,
satisfies the equality
\beq
\chitr \wFz \chitrz =\chitr \tF\chitrz \label{Fequality}
\eeq
which can be verified by computing the matrix elements of both bounded operators
between vectors from $S(\real^s)$, proceeding to configuration space and noting
that the distributions $F$ and $\chi_{-\vz}(\mco_{2(\tr+\teps)})F$ coincide on the 
resulting set of smearing functions. Moreover, we obtain from  (\ref{Fx})
\beqa
|\wFz(\vep)| 
\leq(2\pi)^{-\fr{s}{2}}\int d^sy\, |\chi(\mco_{2(\tr+\teps)})(\vy)| \sup_{|\vec{z}|\leq 2\tr+3\teps} |F(\vec{z}-\vz)|\ph{,}& &\non\\
=c_{s,\tr,\teps}\sup_{|\vec{z}|\leq 2\tr+3\teps} |F(\vec{z}-\vz)|,& &
\eeqa
what concludes the proof of the lemma. \qed\\
After this preparation we set $g=\Lpm e$ in Lemma \ref{harmonic} and undertake the study of the functions
\beq
\real^{s+1}\ni x\to\lan \omm\tilh \Lpm e|U(x)\omm \tilh \Lpm e\ran \label{function}
\eeq
appearing on the right-hand side of  estimate (\ref{harmonice}). We recall from our discussion in Section~2
that $\omm\tilh^{1/2}\Lpm$ are trace-class operators, so $\tilh g$ are in the domain of $\omm$ as
required in Lemma \ref{harmonic}. A link with Lemma \ref{F} is provided by the following identities
\beq
\Lpm=\om^{\mp\half}\chir\om^{\pm\half}\Lpm,  \label{chil}
\eeq
where $\rz$ is the radius of the ball entering into the definition of the
subspaces $\Lpm$. The following result covers the case of translations in space.
\bel \label{mbounds} Assume that $s\geq 3$ and let $e$ be a normalized eigenvector of the operator $T$ corresponding to the eigenvalue $t$. Then there holds 
\begin{enumerate}
\item[(a)] $\lan \om^{-\half}\tilh\Lm e | U(\vz)\om^{-\half}\tilh\Lm e\ran=0$ for $|\vz|>4\rz$,
\item[(b)] $|\lan \om^{-\half}\tilh\Lpm e | U(\vz)\om^{-\half}\tilh\Lpm e\ran|
\leq \fr{c_{s,\rz}t^2}{(|\vz|+1)^{s-2} }$,
\end{enumerate}
where the constant $c_{s,\rz}$ is independent of $\vz$ and $e$.
\eel
\proof
To prove part (a) we set again $\chi_{\vz}(\mco_{\rz})(\vy)=\chi(\mco_{\rz})(\vy-\vz)$ and note that
\beqa
& &\lan \om^{-\half}\tilh\Lm e |U(\vz)\om^{-\half}\tilh\Lm e\ran\non\\
& &\phantom{44444444444}=\lan\om^{-\half}\tilh \Lm e |\chi(\mco_{2\rz})\chi_{\vz}(\mco_{2\rz}) U(\vz)
\om^{-\half}\tilh \Lm e\ran=0,
\eeqa
for $|\vz|>4\rz$, since $\hz\in D(\mco_{\rz})$ and hence $\omm\tilh\Lm e\in [\widetilde{D}(\mco_{2\rz})]$. 
Due to the uniform bound
\beq
|\lan \om^{-\half}\tilh\Lpm e | U(\vz)\om^{-\half}\tilh\Lpm e\ran|\leq 
\|\om^{\alr-\h}\tilh^{1/2}\|^2_{\infty} \lan e |\Tbpm^2 e\ran\leq \|\om^{2\alr-1}\tilh\|_{\infty} t^2,
\label{uniformbound}
\eeq
which involves the parameter $\alr\in [\fr{1}{2},\fr{s-1}{2}[$ from the
definition of the operator $T$, there also follows the ($-$) part of (b). To prove the (+) part we  estimate
\beqa
|\lan\om^{-\half}\tilh\Lp e | U(\vz)\om^{-\half}\tilh\Lp e\ran |
&=&|\lan\tilh\om^\h \Lp e|\chi(\mco_{2\rz})\om^{-2}\chi_{\vz}(\mco_{2\rz})\tilh\omp U(\vz)\Lp e\ran|\non\\
&\leq&t^2\| \om^{2\alr+1} \tilh\|_{\infty}\,  \|\chi(\mco_{2\rz})\om^{-2}\chi_{\vz}(\mco_{2\rz})\|.
\eeqa
Now we are in position to apply  Lemma \ref{F}: We set $\tF(\vep)=|\vep|^{-2}$. Then
\beq
\tF(\vep)^{1/2}=|\vep|^{-1}\theta(-|\vep|+1)+|\vep|^{-1}\theta(|\vep|-1)\in 
L^2(\real^s,d^sp)+L^{\infty}(\real^s,d^sp)
\eeq
and  $F(\vx)=c_s|\vx|^{-(s-2)}$,
where $c_s=\cs$.  We obtain for $|\vz|\geq 6\rz+3\teps$
\beq
\|\chi(\mco_{2\rz})\om^{-2}\chi_{\vz}(\mco_{2\rz})\|\leq  \fr{c_{s,\rz}}{(|\vz|-4\rz-3\teps)^{s-2}}.
\eeq
Making use of the uniform bound (\ref{uniformbound}),
we get the estimate from the statement of the lemma for a suitable constant $c_{s,\rz}$. \qed\\
In order to obtain estimates on  functions (\ref{function}) valid for arbitrary spacelike translations 
$x$ we recall, in a slightly generalized form, the following result from \cite{universal}.
\bel\label{damping} Let $\de>0$. Then there exists some continuous function $f(\om)$ which decreases almost
exponentially, i.e. $\sup_{\om}|f(\om)|e^{|\om|^\kappa}<\infty \textrm{ for any } 0<\kappa<1$,
and which has the property that for any pair of operators
$A$, $B$ such that $\vac$ belongs to their domains and to the domains of their adjoints,
satisfying 
\beq
(\vac| \, [A, e^{itH}Be^{-itH}] \, \vac)=0 \textrm{ for } |t|<\de,
\eeq
there holds the identity $(\vac|AB\vac)=(\vac|Af(\de H)B\vac)+(\vac|Bf(\de H)A\vac)$.
\eel
With the help of the above lemma we prove the desired bounds.
\bel\label{Cook} Assume that $s\geq 3$. Let $e\in L^2(\real^s, d^sp)_1$ satisfy $Te=te$ and $Je=e$.
Then, for any $\eps>0$ and $x\in\real^{s+1}$ s.t. $|\vz|\geq |x^0|$, there hold the estimates
\beq
|\lan\tilh\omm\Lpm e|U(x)\tilh\omm\Lpm e\ran |\leq \fr{c_{s,\rz,\eps}t^2}{(|\vz|-|x^0|+1)^{s-2-\eps}},
\eeq
where the constant $c_{s,\rz,\eps}$ is independent of $x$ and $e$.
\eel
\proof
First, we define the operators $\phi_{+}(e)=a^*(\tilh\Lp e)+a(\tilh\Lp e)$, $\phi_{-}(e)=a^*(i\tilh\Lm e)
+a(i\tilh\Lm e)$ and their
translates $\phi_{\pm}(e)(x)=U(x)\phi_{\pm}(e)U(x)^{-1}$. Since the projections $\Lpm$ and the multiplication
operators $\tilh$ commute with
$J$ and $Je=e$, the operators $\phi_{\pm}(e)$ are just canonical
fields and momenta of the free field theory localized in the double cone of radius $2\rz$ centered at zero.
We assume without loss of generality that $x^0>0$, introduce functions
$\Fpm(\tau)=\lan\tilh\Lpm e|\om^{-1}U(\vz+\tau\te_0)\tilh\Lpm e\ran$ for $0\leq \tau\leq x^0$, where $\te_0$ is the unit vector in the time direction, and consider the derivative
\beq
\bigg|\fr{d\Fpm(\tau)}{d\tau}\bigg|=|(\vac|\phi_{\pm}(e)\phi_{\pm}(e)(\vz+\tau\te_0)\vac)|.
\eeq
We define $\de_{\tau}=|\vz|-\tau-4\rz$ and assume that $\de_{\tau}>0$ for $0\leq \tau\leq x^0$, i.e. $|\vz|-x^0>4\rz$. Then, by locality, $\phi_{\pm}(e)$ and
$\phi_{\pm}(e)(\vz+\tau \te_0)$ satisfy the assumptions of Lemma~\ref{damping} with $\de=\de_{\tau}$. Making 
use of this result, we obtain 
\beqa
\bigg|\fr{d\Fpm(\tau)}{d\tau}\bigg|&=&|\lan\omar\tilh\Lpm e|\om^{2\alr}f(\de_\tau\om)U(\vz+\tau \te_0)\omar\tilh\Lpm e\ran\non\\
&+&\lan\omar\tilh\Lpm e|\om^{2\alr}f(\de_\tau\om) U(-\vz-\tau \te_0)\omar\tilh\Lpm e\ran|\non\\
&\leq& \fr{2}{\de_\tau^{2\alr}} t^2 \| \tilh\|_{\infty}\, \sup_{\om\geq 0}|\om^{2\alr}f(\om)|. \label{derivative}
\eeqa
Next, we set $\alr=\fr{s-1-\eps}{2}$ for $0<\eps<1$ and arrive at the following estimate
\beqa
|\lan \omm\tilh\Lpm e|U(x)\omm\tilh\Lpm e\ran |=|\Fpm(x^0)|&\leq& |\Fpm(0)|+\int_0^{x^0}d\tau \bigg|\fr{d\Fpm(\tau)}{d\tau}\bigg|\non\\
&\leq& \fr{c_{s,\rz,\eps}t^2}{(|\vz|-x^0-4\rz)^{s-2-\eps} },\label{Cookmethod}
\eeqa
where in the last step we applied Lemma \ref{mbounds} and estimate (\ref{derivative}).
Since the left-hand side of relation (\ref{Cookmethod}) satisfies a uniform bound analogous to (\ref{uniformbound}),
we obtain the estimate in the statement of the lemma. \qed\\
Now we are ready to prove the required bounds on the norms of the functionals~$\Smn$.
\bep\label{semibound} Given a family of points $\x\in\real^{s+1}$ we define 
$\de(\vxb)=\inf_{i\neq j}(|\vx_i-\vx_j|-|x_i^0-x_j^0|)$. For $s\geq 3$, $\de(\vxb)\geq 0$ and any $\eps>0$ the functionals
$\Smn$ satisfy the bound
\beqa
\|\Smn\|_{\x}^2
&\leq& 16 c_{s,\rz,\eps}\cErs E^{|\mub|+|\nub|}t^{2(\mub+\nub)}\bigg\{1+\fr{N-1}{(\de(\vxb)+1)^{s-2-\eps}}\bigg\},\,\,
\eeqa
where the constant $c_{s,\rz,\eps}$ appeared in Lemma~\ref{Cook}.
\eep
\proof Making use of the fact that $S_{0,0}=0$, we can assume without loss of generality
that $\nub\neq 0$ and decompose it into two pairs of multiindices $\nub=\nuh+\nuc$ in such a 
way that $|\nuc|=1$. Proceeding as in the proof of Proposition 4.4 in \cite{Dyb} (formulas
(4.12) and (4.13)) we obtain the bound
\beq
\|\Smn\|^2_{\x}\leq 16E^{|\mub|+|\nuh|}t^{2(\mub+\nuh)}\|P_E\sum_{k=1}^N\big(a^*(\LL e)^{\nuc}a(\LL e)^{\nuc}\big)(x_k)P_E\|.\label{aux2}
\eeq
From Lemmas \ref{harmonic} and  \ref{Cook} we get 
\beqa
& &\|P_E\sum_{k=1}^N\big(a^*(\LL e)^{\nuc}a(\LL e)^{\nuc}\big)(x_k)P_E\|
\leq 
E\cErs\big\{\|\tilh\omm(\LL e)^{\nuc}\|^2 \non\\
& &\phantom{4444444444444}+(N-1)\sup_{i\neq j}|\lan\tilh\omm(\LL e)^{\nuc}| U(x_i-x_j)\tilh\omm(\LL e)^{\nuc}\ran| \big\}\non\\
& &\phantom{4444444444444}\leq c_{s,\rz,\eps} \cErs E t^{2\nuc}\bigg\{1+\fr{N-1}{(\de(\vxb)+1)^{s-2-\eps}}\bigg\}.
\label{collect}
\eeqa
Substituting inequality
(\ref{collect}) into  formula (\ref{aux2}),
we obtain the estimate in the statement of the proposition. \qed\\
We note that the bound from Proposition \ref{semibound} has a similar structure to estimate~(\ref{Sestimate})
for the ordinary norms of $\Smn$. Therefore, making use of formulas (\ref{start}) and (\ref{traces}), we obtain
\beqa
& &\|\PiE\|_{p,\x}\non\\ 
& &\phantom{44}\leq 4c_{s,\rz,\eps}^{1/2} \cE \bigg(\sum_{k=0}^\infty \fr{(2^5E)^{\half pk}\|T^p\|_1^k}{(k!)^{\half p}} \bigg)^{\fr{4}{p}}
\bigg\{1+\fr{N-1}{(\de(\vxb)+1)^{s-2-\eps}}\bigg\}^\h.\,\,\,\,\,\,\,\,
\eeqa
It follows that  $\limsup_{\de(\vxb)\to\infty}\|\PiE\|_{p,\x}$ satisfies a bound which is  
independent of $N$. Consequently, we get
\bet  Condition $\Nnat$ holds in massless scalar free field theory in $s\geq 3$ dimensional space.
\eet

\section{Conclusions}

In this work we verified the sharpened nuclearity condition $\Nnat$ in massless free field theory
in spacetime of physical or higher dimension. This criterion guarantees the uniqueness of the vacuum 
state in the energy-connected component of 
the state space,  in agreement with physical observations \cite{Dyb}. Nevertheless,
it turns out to be consistent with a degenerate vacuum  structure: 
Recall that massless free field theory has a spontaneously broken gauge symmetry $\real\ni\la\to\be_{\la}$,
corresponding to a shift of the pointlike localized field by a constant, which is defined on Weyl operators~by
\beq
\be_{\la}(W(f))=e^{i\la(\wt{\om^{1/2} f})(0)}W(f). \label{gauge}
\eeq
This group of transformations gives rise to a family of pure, regular vacuum states
\beq
\om_0^{(\la)}(W(f))=e^{i\la(\wt{\om^{1/2} f})(0)}\om_0(W(f))
\eeq
whose energy-connected components are, in fact, disjoint subsets of the state 
space for $s\geq 3$ \cite{BWa}. This is no longer true for $s=2$  in which case Condition $\Nnat$,
as well as the weaker Condition $\Ns$, does not hold due to singular infrared properties of this theory \cite{BP}.

The methods developed in the present Letter are relevant to harmonic analysis
of local operators $A\in\mfa(\mco)$. We recall that in any  relativistic quantum 
field theory there holds the bound \cite{Buch3}
\beq
\sup_{\fun\in\traceEB}\int d^sp|\vep|^{s+1+\eps}|\fun(\wt{A}(\vep))|^2<\infty, \label{ha}
\eeq
for any $\eps>0$, where  $\wt{A}(\vep)$ is the Fourier transform of $A(\vx)$. Since the mollifier
$|\vep|^{s+1+\eps}$ suppresses the contributions to $\fun(\wt{A}(\vep))$ with small momentum  transfer,
which become relevant at asymptotic times \cite{AH,Porr1,Porr2},
we are interested in the minimal power of $|\vep|$  for which the bound (\ref{ha}) is still
valid. Making use of an improved variant of an improved variant of Lemma \ref{harmonic}, 
one can show that for $s\geq 3$ there holds in massless free field theory
\beq
\sup_{\fun\in\traceEB }\int d^sp|\vep|^{2}|\fun(\wt{A}(\vep))|^2<\infty. \label{ha1}
\eeq
With the help of a suitable sequence of functionals $\fun_n\in\traceEB$, involving arbitrarily
large number of particles, it can be verified that the power of the mollifier $|\vep|^2$ cannot be 
further reduced on the whole local algebra $\mfa(\mco)$ in this model.
However, making use of the more refined expansion of the map $\PiE$ into rank-one mappings, developed in \cite{Bos3},
one can construct a subspace \emph{of finite co-dimension} in $\mfa(\mco)$ on which there holds the bound
\beq
\sup_{\fun\in\traceEB}\int d^sp|\fun(\wt{A}(\vep))|^2<\infty, \label{ha2}
\eeq
familiar from massive free field theory \cite{Dyb}. This subspace contains, in particular, the elements
of the fixed-point subalgebra of $\la\to\beta_{\la}$ whose vacuum expectation
values vanish. These results, whose detailed proofs will be presented elsewhere, demonstrate the
utility of the phase space methods in the development of a more detailed harmonic analysis of automorphism groups 
\cite{Arveson}.

\bigskip

\noindent{\bf Acknowledgements:}
I would like to thank Prof. D. Buchholz for his continuing advice and encouragement
in the course of this work. Financial support from Deutsche Forschungsgemeinschaft is 
gratefully acknowledged.

\end{document}